# Incorporating Si into Sb$_2$Se$_3$: Tailoring Optical Phase Change Materials via Nanocomposites

Chih-Yu Lee[1], Yi-Siou Huang[1,2], Felix Adams[1], Chuanyu Lian[1,2], Hongyi Sun[1,2], Jie Zhao[3], Zichao Ye[3], Nathan Youngblood[4], Juejun Hu[5], Leslie H Allen[3], Yifei Mo[1], Ichiro Takeuchi[1,6*], Carlos A Rios Ocampo[1,2*]

[1] Department of Materials Science and Engineering, University of Maryland, College Park, MD 20742, USA
[2] Institute of Research in Electronics and Applied Physics, University of Maryland, College Park, Maryland, USA
[3] Department of Materials Science and Engineering, University of Illinois at Urbana-Champaign, Urbana, Illinois, USA
[4] Electrical and Computer Engineering Department, The University of Pittsburgh, Pittsburgh, PA, USA
[5] Department of Materials Science & Engineering, MIT, Cambridge, MA, USA
[6] Maryland Quantum Materials Center, University of Maryland, College Park, Maryland, USA
 *Corresponding authors: takeuchi@umd.edu, riosc@umd.edu

## Abstract

Chalcogenide-based optical phase change materials (OPCMs) exhibit a large contrast in refractive index when reversibly switched between their stable amorphous and crystalline states. OPCMs have rapidly gained attention due to their versatility as nonvolatile amplitude or phase modulators in various photonic devices. However, open challenges remain, such as achieving reliable response and transparency spanning into the visible spectrum, a combination of properties in which current broadband OPCMs (e.g., Ge$_2$Sb$_2$Se$_4$Te$_1$, Sb$_2$Se$_3$, or Sb$_2$S$_3$) fall short. Discovering novel materials or engineering existing ones is, therefore, crucial in extending the application scope of OPCMs. Here, we use magnetron co-sputtering to study the effects of Si doping into Sb$_2$Se$_3$. We employ ellipsometry, X-ray diffraction, Raman spectroscopy, and scanning and transmission electron microscopy to investigate the effects of Si doping on the optical properties and crystal structure and compare these results with those from first principles calculations. Moreover, we study the crystallization and melt-quenching of thin films via nano-differential scanning calorimetry (NanoDSC). Our experiments demonstrate that 20% Si doping increases the transparency window in both states, specifically to 800 nm (1.55 eV) in the amorphous phase, while reducing power consumption by lowering the melting temperature. However, this reduction comes at the cost of reducing the refractive index contrast between states and slowing the kinetics of the phase transition. Moreover, we demonstrate that Si-Sb$_2$Se$_3$ solid solution can be reversibly switched between the amorphous and crystalline states using electro-thermal switching in photonic integrated devices, ~100 μm$^2$ microheaters, and in the NanoDSC sensor itself. We observe a phase separation in recrystallized films, which does not hinder the reversible switching associated with Sb$_2$Se$_3$ and allows an effective medium response in the optical properties; thus, opening a route towards tailoring optical properties of phase change by engineering nanocomposites.

# 1. Introduction

Chalcogenide-based optical phase change materials (OPCMs) have garnered attention in a wide range of applications, including photonic circuits[1–4], structural color[5], optical metamaterials[6–8], polariton nanophotonics[9], and others. These applications leverage OPCM's nonvolatile optical responses and exploit the contrast in their refractive index and extinction coefficient. Finding novel OPCMs with a large refractive index, vanishing extinction coefficient, low power consumption, transparency window spanning into the visible spectrum, and high switching speed has therefore become essential for advancing these emerging fields.[10]

Low-loss, wide-bandgap OPCMs can be engineered by introducing light chalcogens into conventional Te-based phase change materials (PCMs). The elemental substitution of tellurium with selenium in $Ge_2Sb_2Te_5$, for instance, significantly increases their optical bandgap, from 0.62 eV ($Ge_2Sb_2Te_5$) to 1.64 eV ($Ge_2Sb_2Se_5$). Hence, controlling the Se/Te content in the alloys enables tailoring the materials' optical properties.[11] $Ge_2Sb_2Se_4Te_1$[12], an intermediate alloy in the Te substitution with Se, is a well-known example of an OPCM with optimized properties, combining broadband transparency (1–18.5μm in the amorphous and ~2.4-18.5 μm in the crystalline state), large optical contrast (contrast in refractive index, $\Delta n$ = 1.5), and improved thermal stability (crystallization temperature, $T_c$~300°C). As for binary compounds, $Sb_2Se_3$, previously studied in solar cell absorbers[13] and highly sensitive photodetectors[14], has become a popular OPCM for pure phase modulation in the telecommunication C-band (1550 nm), since it exhibits vanishing losses (k < $10^{-5}$) in both the amorphous and crystalline states.[15] $Sb_2Se_3$ undergoes congruent melting, making it a robust material with demonstrated millions of switching cycles.[16,17] While $Sb_2Se_3$ features an absorption edge close to ~900 nm, another OPCM, $Sb_2S_3$, stands out with even wider transparency down to ~600 nm in the amorphous state. Both $Sb_2Se_3$ and $Sb_2S_3$ are promising candidates for photonic applications; however, $Sb_2Se_3$ achieves an optimal balance between thermal stability and optical contrast and has proven to be more chemically inert and durable.[18] $Sb_2Se_3$, however, could benefit from further materials engineering to enhance its amorphous stability ($T_c$~200°C) and transparency in the visible range.

While OPCMs can be tailored via modulating the number of electrons shared between adjacent atoms, i.e., chemical bonding[19], discovering new OPCMs using first-principles has proved difficult, particularly in the amorphous state[20]. Instead, incorporating dopants into a matrix of existing OPCMs via combinatorial deposition is a faster route to obtaining optimal materials, following a strategy that has been exploited widely in electrical phase change materials. For example, N[21], O[22] and C[23] are exploited as dielectric dopants, and Sc[24], Mn[25], and Cr[26] are considered as metal dopants to PCMs. In some previous studies, doping increases crystallization temperature and resistivity in amorphous and/or crystalline states, such as N-doped $Ge_2Sb_2Se_5$[21] and O-doped Sb[27], due to the increased number of scattering sites and grain boundaries. Other examples include doping $SiO_2$ into an Sb matrix to create a nanocomposite with enhanced thermal properties[28] and doping metal oxides into Sb.[29] In both cases, Sb undergoes a phase transition, while the oxides act as a three-dimensional confinement for adjacent regions.

Previous studies on doped $Sb_2Se_3$ focus on photovoltaic solar cells, such as Sn-doped $Sb_2Se_3$[30] and Cu-doped $Sb_2Se_3$[31]. Hu *et al.* showed Si-doped Sb₂Se (not $Sb_2Se_3$) as a promising PCM because of its higher crystallization temperature, larger crystallization activation energy, and better data retention ability.[32] Liu *et al.* demonstrated Si-doped Sb for higher stability in ultrafast operation.[33] However, there is a paucity of studies on the effects of dopants on the optical properties of OPCMs—specifically, their refractive index and extinction coefficient, as well as their phase change behavior. Here, we employ combinatorial and co-sputtering methods, shown in **Figure 1a**, to systematically investigate the effects of 0-20% Si doping in $Sb_2Se_3$ thin films and provide optical, structural, and thermal characterization data. We also perform Density Functional Theory (DFT) calculations on Si-doped $Sb_2Se_3$. We investigate the thermodynamics of Si-doped $Sb_2Se_3$ materials using ultrafast nano-differential scanning calorimetry (NanoDSC), which directly measures the quantitative thermal metrics (temperature, enthalpy, and heat capacity) by cycling OPCMs between amorphous and crystalline states. To further demonstrate reversible switching, we integrate Si-doped $Sb_2Se_3$ into photonic integrated circuits (PICs) with embedded microheaters for electro-thermal dynamic switching. By measuring the phase shifts upon OPCM switching, the PIC enables, in addition, measuring the OPCM's optical properties. Lastly, we performed reflectance measurements directly on the binary compound $SiSe_2$ - $Sb_2Se_3$, deposited on large microheaters.

## 2. Materials characterization

We start by characterizing thin films with binary composition spreads from Si and $Sb_2Se_3$ targets on 400 μm-thick, 3-inch sapphire wafers using an ultrahigh-vacuum (base pressure: $2 \times 10^{-8}$ Torr) magnetron sputtering system. We deposited both continuous gradients and $3.5 \times 3.5$ mm$^2$ pads (see **Figure 1a**) by using a Si mask and proceeded to perform optical, structural, and thermal characterization. See Experimental Section for more details about sample preparation.

### 2.1 Optical properties

We used the Tauc-Lorentz model to fit the complex refractive index of the different compositions and derive the optical band gaps from the dispersion model, as shown in **Figure 1b-d**. The refractive index ($n$), extinction coefficient ($k$) and band gap ($E_g$) of pure $Sb_2Se_3$ are comparable to those reported in literature.[34,35] Note that optical constants vary slightly across different synthesis methods owing to variations in film morphology, roughness, density, and microstructure. Hence, we considered undoped $Sb_2Se_3$ as the baseline for all comparisons. We prepared Si and $Sb_2Se_3$ compositional combinations in both patterned spreads and in continuous spreads to avoid edge-related artifacts in the ellipsometry characterization (see Supplementary **Figures S1 and S2**). For the as-deposited amorphous film, the $E_g$ changes from 1.37 eV to 1.50 eV as the Si concentration increases from 0% to 20%, as shown in **Figure 1b**. The $E_g$ first reduces slightly as Si reaches 1.5% and remains at ~1.30 eV until Si attains 13.4%. However, the bandgap climbs drastically after Si doping exceeds 15%. We note that the error margin of dispersive model fitting based on Tauc-Lorentz dispersion oscillators ranges from 0.002 to 0.008 eV; therefore, the bandgap fitting is considered reliable. The bandgap narrowing at low doping concentration can be attributed to band renormalization with impurities or band tailing, which is often observed as Urbach tailing.[36,37] **Figure 1c** shows the refractive indices as a function of wavelength. While the absorption edge expands to 828 nm as Si increases, the value of $n$ decreases. This is expected given Kramers-Kronig relations, with high $n$ associated with low $E_g$, influenced also by several factors, such as electronic structure, phonon interaction, and band structure, that affect the relationship between them.[38,39] Given that the phase transition temperature for each composition was unknown, we prepared samples annealed at 200 °C, 240 °C, and 350 °C, respectively. We see, in **Figure 1b**, that the $E_g$ of most compositions drops to 1.1 eV after 200 °C annealing for over 20 minutes, which may indicate that the materials are fully crystallized. However, compositions with high Si% (>10%) required a higher temperature to achieve a stable $E_g$, which for 20% Si decreases to ~1.2 eV after annealing at 350 °C. The missing data points for 200 °C and 240 °C annealed samples exhibited abnormally low $E_g$ values due to unreliable fitting arising from the edge effect in patterned spreads. The associated refractive indices are plotted in **Figure 1d**. Despite displaying a larger bandgap and contrast in optical properties, the crystallization of thin films proved challenging for higher Si compositions, which triggered our study on their phase transformations and associated thermodynamics/kinetics.

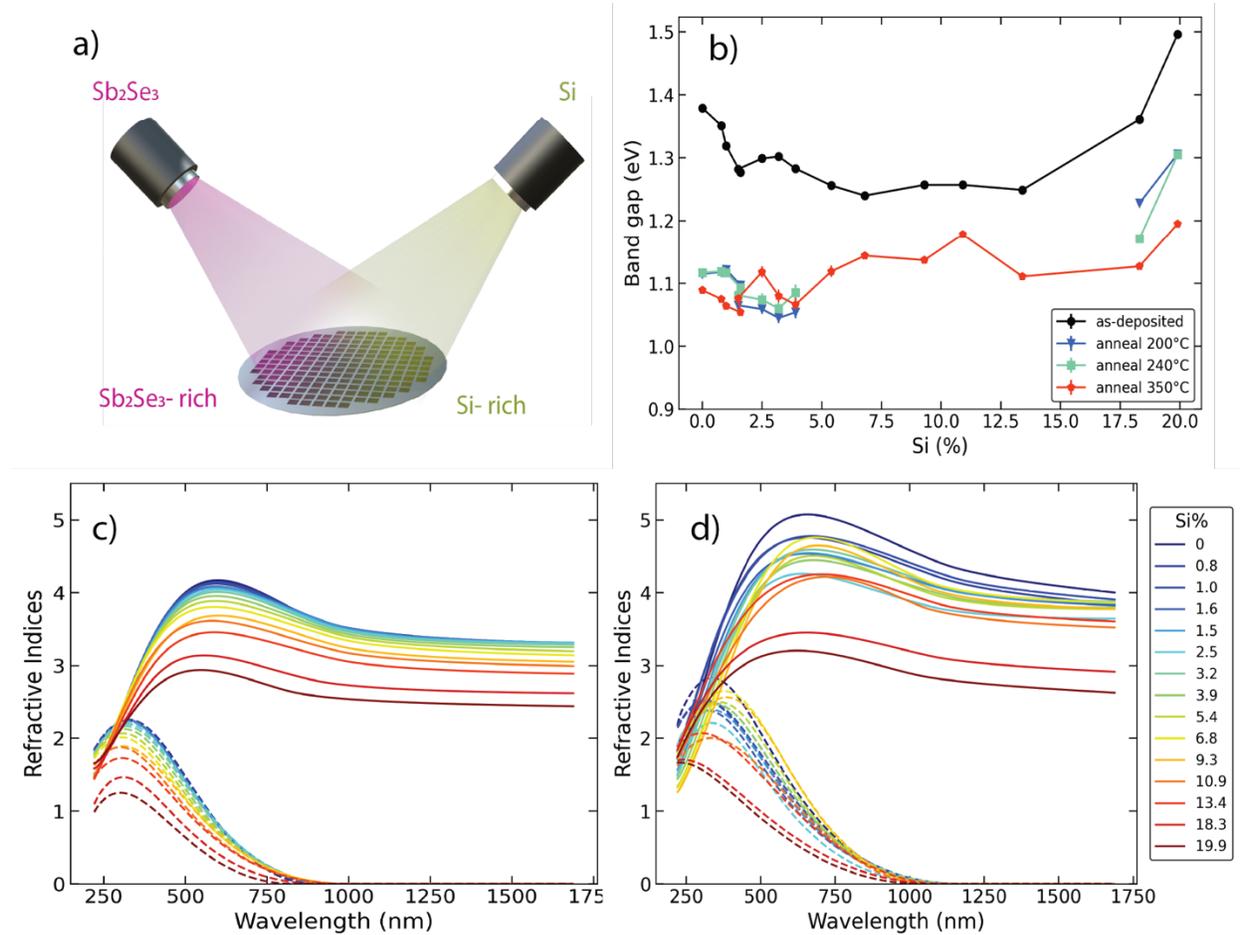

**Figure 1. Bandgap and refractive index characterization**. (a) Schematic of co-sputtering. (b) Optical bandgap extracted from ellipsometry measurement. The data was collected from different spreads (as deposited, annealed at 200 °C, 240 °C, and 350 °C) across 15 compositions. (c) and (d) are wavelength-dependent refractive indices of different compositions measured as deposited and 350°C annealed, respectively. The solid lines correspond to the refractive index ($n$) and dashed lines to the extinction coefficient ($k$).

## 2.2 Density Functional Theory (DFT) calculations

We performed DFT calculations to investigate the effect of Si doping in crystalline $Sb_2Se_3$ from first principles. The computational models are based on single-phase $Sb_2Se_3$ with Si and vacancies on Sb sites, resulting in different compositions but similar Si concentrations as observed in our experiments (**Figure 2a**)—see the Experimental Section for more details and Supplementary Figure S3 for the crystal structures. **Figure 2b** shows the DFT predicted bandgap as a function of Si concentration, which resembles the experimental results in Figure 1b, but the predicted bandgaps are lower, which could be attributed to a well-known systematic error in DFT calculations[40]. While there is no significant change in the band gap at low Si concentrations, the bandgap increases by about 30% at the highest concentration calculated (15.8 at%).

We note that these results assume that the material remains in a single phase, i.e., no crystal phase segregation.

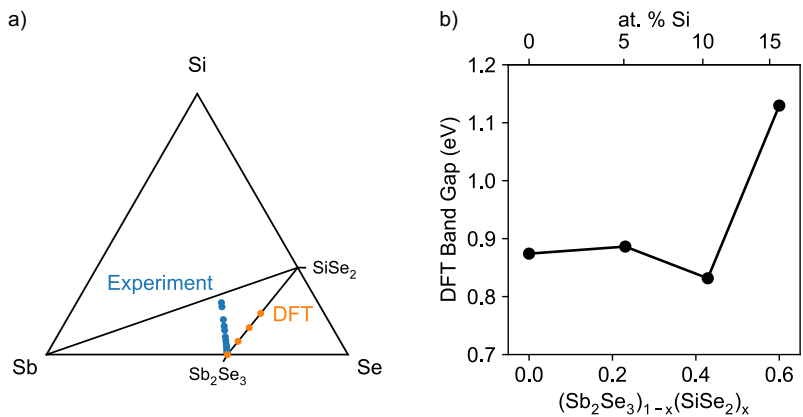

**Figure 2. The effects of Si substitution into Sb$_2$Se$_3$ as predicted by DFT.** (a) The composition of the DFT supercells compared to the experimentally measured compositions. (b) The band gap predicted by DFT as a function of Si concentration. The crystal structures are shown in Supplementary Figure S3.

## 2.3 X-ray Diffraction (XRD)

We measured temperature-dependent XRD on two samples: pure Sb$_2$Se$_3$ and 20% Si-doped Sb$_2$Se$_3$ to obtain structural information on as-deposited, crystalline, molten, and recrystallized states. In **Figure 3a**, the diffraction peaks for pure Sb$_2$Se$_3$ appear at 200°C and disappear after 600 °C, which is consistent with the crystallization and melting temperatures, $T_c$ and $T_m$, reported in the literature.[41] The XRD patterns are identified as orthorhombic Sb$_2$Se$_3$, again matching previous studies. **Figure 3b** demonstrates that 20% Si-doped Sb$_2$Se$_3$ crystallizes at 400 °C, and surprisingly, it melts at 500 °C, indicating a relatively narrow operational window for achieving crystallization on a device. The preferred orientations are also slightly different from pure Sb$_2$Se$_3$. For example, the (020) and (120) orientations are not obvious, whereas the (130) orientation is preferred; no additional peaks were observed. Except for lower crystallinity and smaller grain size based on the Scherrer equation, the long-range order crystal structure is similar to that of pure Sb$_2$Se$_3$. They both exhibit a shift towards higher 2θ in recrystallization compared to the first crystallization, which might indicate the compressive force applied on the films by the capping layer while the crystal grows. We do not observe a second phase with a different crystal structure. The temperature-dependent XRD results align well with those of the hot-plate annealed samples in Figure 1, in that Si-doped samples are more challenging to crystallize. With these results, we confirm that Si dopants suppress the crystallization of amorphous films, similar to previously reported results for Si-doped Sb$_2$Te$_3$.[43]

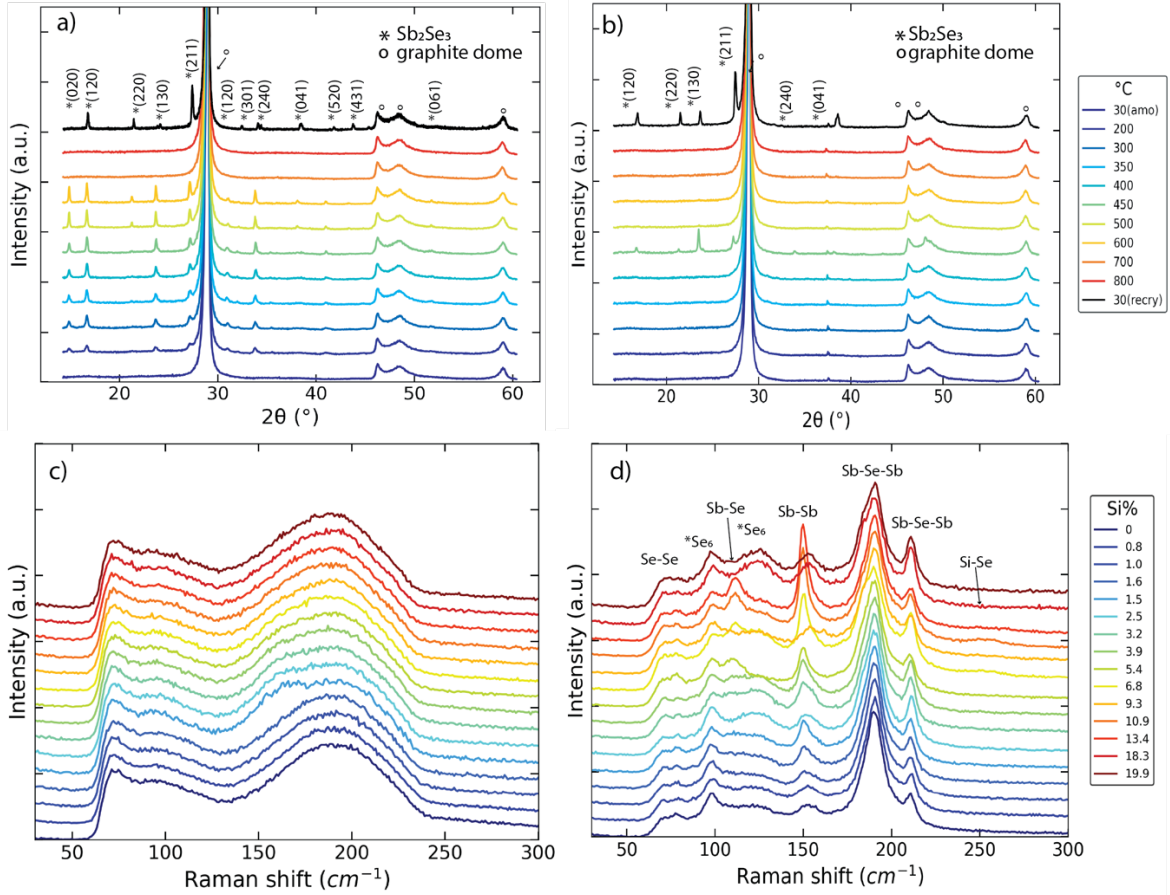

**Figure 3. Temperature-dependent XRD**. (a) $Sb_2Se_3$ and (b) 20% Si-doped $Sb_2Se_3$ from 30 °C amorphous (amo) to 800 °C, and then quench down to room temperature, 30 °C (recry). Raman spectroscopy for 15 compositions measured (c) as-deposited and (d) after 400 °C annealing for over 20 minutes.

## 2.4 Raman Spectroscopy

We employed Raman spectroscopy to probe the bonding interactions in the short-range order. This technique also enables us to monitor changes in bonding characteristics following the annealing process across various Si concentrations. In **Figure 3c**, all compositions of as-deposited Si-doped $Sb_2Se_3$ demonstrate a similar Raman spectrum, exhibiting a prominent broad peak centered at approximately 190 cm$^{-1}$, which corresponds to the characteristic signature of amorphous $Sb_2Se_3$ [15]. No Si-Si peak ~500 cm$^{-1}$ was observed in any of the samples, which is why we focus on analyzing Raman signatures below 300 cm$^{-1}$. Following the annealing process at 400 °C for 20 minutes on a hotplate, the previously observed broad peak near 190 cm$^{-1}$ resolved into two distinct, sharper peaks at approximately 190 cm$^{-1}$ and 210 cm$^{-1}$. This transition indicates an increase in $Sb_2Se_3$ crystallinity. The resulting peaks are attributed to the Sb−Se−Sb bending vibrations characteristic of the $Sb_2Se_3$ orthorhombic structure.[13,42] Additional Raman peaks are observed after the annealing process. The peaks at 110 and 150 cm$^{-1}$ indicate the presence of Sb-Sb bonding.[44,45] The peaks around 100 and 130 cm$^{-1}$ are attributed to the $Se_6$ rhombohedral ring structure, while the features near 80 cm$^{-1}$ are linked to Se-Se bonding contributions. Notably, the Raman peak observed at around 250 cm$^{-1}$ appears in the sample regions with relatively high silicon concentrations and lower $Sb_2Se_3$ concentrations. Therefore, we propose that this peak most likely originates from Si-Se interactions, which we observe more clearly later in Section 3 [47,48]. It is noticeable that the Si-Se resonance bond appears at 6.8% and 10.9% of Si doping, which implies the potential substitution of Si into Sb sites, consistent with DFT calculations, suggesting the formation of SiSe$_2$, the stable Si-Se glass at ambient conditions. This resonance becomes negligible in the highly Si-doped region (>10%), likely due to disorder (i.e., the sample did not fully crystallize after this annealing), indicating that potential Si-Se bonds are still absent.

## 2.5 Nanocalorimetry and TEM

Nano-differential scanning calorimetry, or NanoDSC,[49–51] is a technique that uses a microscale device for thin-film DSC characterization, meeting industry standards with a sensitivity of 1 Å and a scanning rate up to $3 \times 10^6$ K/s. In addition to thermal characterization, we utilize the NanoDSC device to cycle the OPCMs reversibly, simulating a real OPCM device, which alleviates concerns about the transferability of results from characterization systems to their practical application.

To further investigate the thermodynamic effects of Si doping on the phase transition, we use co-sputtering on NanoDSC devices, as shown in **Figures 4a and 4b**. In **Figure 4c,** The $C_p(T)$ of the pre-scanned (the as-deposited sample melted by a scan to 650 °C) 10% Si-doped sample shows a reproducible crystallization-melt similar to the undoped sample, with a $T_c$ roughly 10 °C higher. From the kinetic study in **Figure 4e**, we find that the activation energy of 10% Si-doped materials is lower (0.9 eV) than that of undoped materials (1.2 eV). This activation energy is derived from the Kissinger analysis of the crystallization curve, as described in our prior work [52]. However, the first $C_p(T)$ scan of the as-deposited 20% Si-doped sample, in **Figure 4d**, shows a significantly reduced melting temperature ($T_m$=490 °C), which is 120 °C lower than that of pure Sb$_2$Se$_3$. These results agree with our XRD findings (crystalline peaks disappear at lower temperatures in **Figure 3b**) and the lower switching power observed in PIC platforms (see Section 3.1). Additionally, both the melting and crystallization enthalpy ($C_p(T)$ peak integral area) are much shallower than in undoped Sb$_2$Se$_3$, indicating that only part of the film undergoes the first crystallization-melting-quench cycle while the rest remains amorphous.

For TEM characterization, we applied electrical annealing (calorimetric pulses to 450 °C with a heating rate of 120,000 K/s) to the 20% Si-doped sample after the first melt-quench cycle to maximize its crystallinity, mimicking crystallization pulses in a real device (see **Figure 4f)**. TEM reveals two phases, (g) glassy and (h) polycrystalline, in the recrystallized films. This indicates phase segregation after the first cycle. The (g) phase shows minor dewetting with irregular morphology, and diffusion rings are observed in TEM diffraction patterns (**Figure 4g: 1**). The (h) phase has a more defined round shape, confirmed as grains with different orientations (**Figure 4g: 2-4).** Since the phase is amorphous, there are no crystalline peaks in the XRD patterns of the recrystallized 20% Si-doped sample, as shown in **Figure 3b**. We hypothesize that the (g) phase is associated with Si-rich composition and the formation of SiSe$_2$, supported by a qualitative EDS study, shown in Supplementary Figure S5. However, due to the complexity of the multi-layered device structure and nanoscale features, it was challenging to accurately probe nanocomposites and to quantitatively isolate Si from the background (SiN$_x$ membrane and Si wafer).

We further performed $C_p(T)$ scans on the phase-separated samples in **Figure 4g**. The subsequent scans show reproducible cycling with a much higher $T_m$ and $T_c$ (2$^{nd}$ and 3$^{rd}$ scans) than the first scan in **Figure 4d**. The temperatures are, in fact, close to that of pure Sb$_2$Se$_3$. This indicates that the phase (β) that participates in the subsequent melt-quench cycle is Sb$_2$Se$_3$ with a low concentration of Si. We note that the glass transition peaks, labeled as $T_g$ in **Figure 4f**, are more pronounced compared to those in **Figure 4e**, indicating the existence of an excess amount of glassy phase in the sample. Also, the observation of melting peaks with lower intensities might suggest that only a portion of the material is melting. For the phase-separated 20% Si-doped sample, the cycling portion shows higher activation energy (1.6 eV) compared to pure $Sb_2Se_3$, as summarized in Table 1. The NanoDSC results suggest, therefore, that the first melting of the as-deposited 20% Si-doped $Sb_2Se_3$ sample is much lower (490 °C) than pure $Sb_2Se_3$. However, after rapid heating, the sample phase-separated into α (glassy, hard to crystallize, Si-rich) and β (easy to crystallize, Si-poor), forming a nanocomposite.

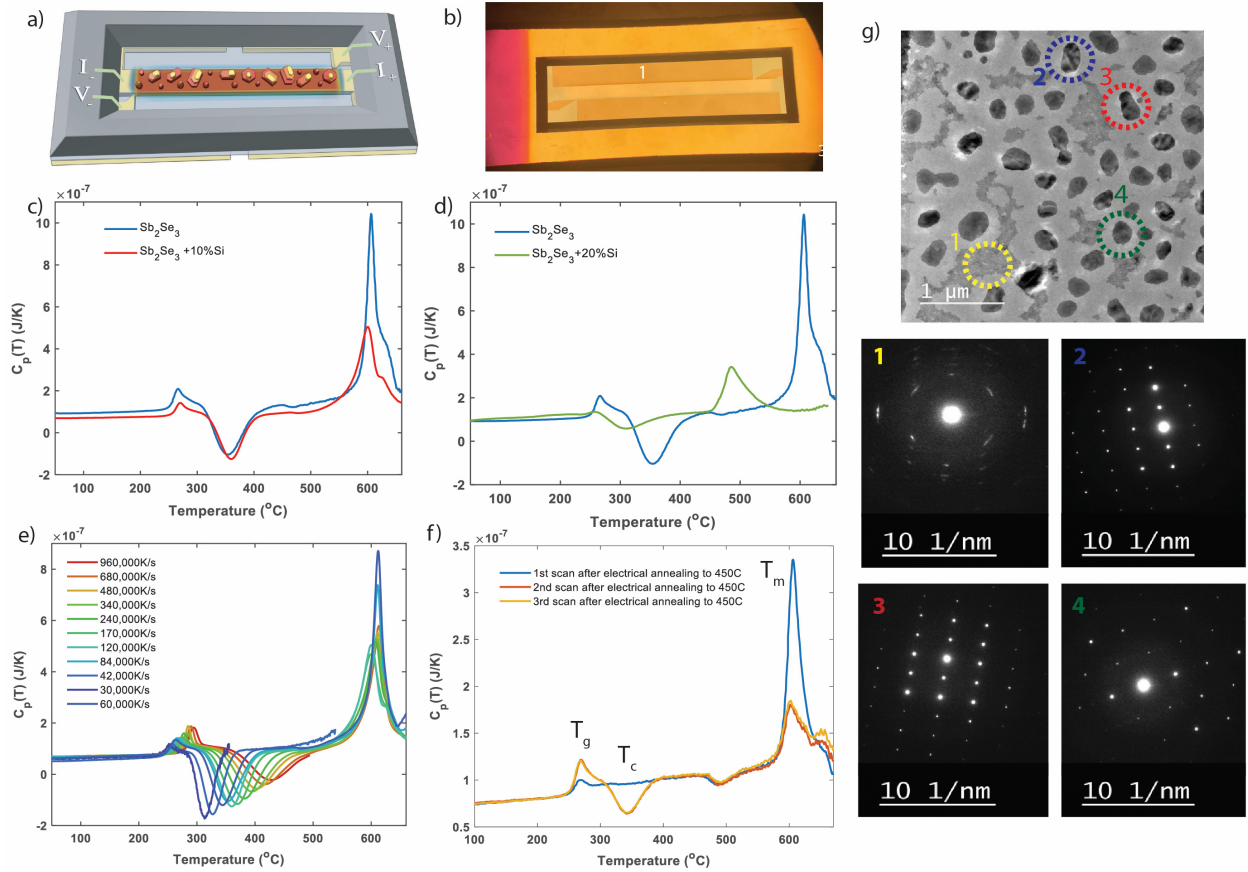

**Figure 4. Reversible switching of Si-doped $Sb_2Se_3$ on NanoDSC.** (a) Schematic of NanoDSC device. (b) Top-view image of NanoDSC. (c) $C_p(T)$ of pre-melted 10%- Si doped sample with a scanning rate of 120,000 K/s. (d) $C_p(T)$ for the first scan of 20%- Si-doped sample. For (c) and (d), $C_p(T)$ of pure $Sb_2Se_3$ from our other work (unpublished) is included for comparison. (e) and (f) are NanoDSC scans at 120000 K/s of the 10%- Si-doped and 20%- Si-doped samples, respectively. (g) TEM images and selected area diffraction patterns for 20%- Si-doped sample after electrical annealing. The number and color in (g) match the TEM images and selected area electron diffractions.

*Table 1.* Thermal properties of Si-doped $Sb_2Se_3$ materials measured through NanoDSC. Activation energy is extracted from kinetic study (scanning rate dependent). Crystallization temperature is based on data collected at 120000 K/s.

| $Si_x(Sb_2Se_3)_{1-x}$ | 0% | 10% | 20% (as-deposited scan) | 20% (recrystallization scans) |
|---|---|---|---|---|
| $E_a$ (eV) | 1.2 | 0.9 | N/A | 1.6 |
| $T_c$ (°C) | 351 | 360 | 310 | 345 |
| $T_m$ (°C) | 606-610 | 598-608 | 484 | 605 |

## 3. Device integration

We now demonstrate that Si- and SiSe$_2$-doped Sb$_2$Se$_3$ can be reversibly switched using microheaters for integrated and free-space optical applications—a fundamental test that every OPCM should pass. Even when forming nanocomposites, Sb$_2$Se$_3$ is actively switching while Si and SiSe$_2$ form stable amorphous domains, resulting in tunable effective medium optical properties.

### 3.1 Reversible switching in photonic integrated circuits

To further investigate the thermal and optical properties of Si-doped Sb$_2$Se$_3$, we integrated 20% Si-doped Sb$_2$Se$_3$ into ring resonators on a silicon-on-insulator platform and performed electro-thermally driven reversible switching. **Figure 5a** displays the device layout on a partially foundry-fabricated chip[53]. Silicon serves both as the waveguide and as a resistive heater in regions with n$^+$ doping, enabling the switching of OPCMs through indirect resistive heating, as shown in **Figure 5b and c**. Supplementary Figures S4, S5, and S6 show an SEM image and the EDS compositional mapping of the device, respectively. We applied a long, low-power pulse for crystallization and a short, high-power pulse for amorphization, as summarized in **Table 2**. A 3 V/50 µs pulse is sufficient to crystallize pure Sb$_2$Se$_3$; however, surprisingly, only 1.85 V is required to achieve crystallization in the 20% of Si-doped sample. Yet, this 30 ms pulse is significantly longer than the undoped counterparts, which suggests slower kinetics in crystallization. While a 7V/ 500ns pulse is needed for amorphization in pure Sb$_2$Se$_3$, a 4.7V/ 1µs pulse is sufficient to re-amorphize 20% Si-doped sample. Remarkably, only ~150 mW is required, compared to the ~250 mW of pure Sb$_2$Se$_3$;[54] however, considerable energy consumption is due to the longer pulse. The evidence of reduced switching power is aligned with the lower $T_m$ observed from temperature-dependent XRD and the first scan on NanoDSC shown above. Using the parameters listed in Table 2, we successfully switched the devices for three cycles. The refractive index contrast in OPCM modulates the effective index of the ring, resulting in a phase shift and thus, a change in the device's resonance wavelengths that we track experimentally, as shown in **Figure 5d**. We were unable to replicate the amount of redshift during recrystallization and the blue shift in re-amorphization observed in the first cycle, which we attribute to phase segregation (i.e., nanocomposite formation) occurring after the first full switching, as discussed in Section 2.5. The total phase shift in 20% Si-doped (0.2π) is smaller than that of undoped Sb$_2$Se$_3$ (~π) for a 12 µm long cell, which is expected, given the smaller contrast in $n$ upon reversible switching (see **Figure 1b**).

By tracking the resonant wavelength shift ($\lambda_{res}$), we calculated the refractive index contrast of the material on the PIC ($\Delta n_{eff}$) using:

$$\Delta n_{eff} = \frac{\lambda_{res} \times \Delta\varphi}{2\pi \times L},$$

where $\Delta\varphi$ denotes phase shift and is given by $\Delta\varphi = \frac{2\pi \times \Delta\lambda_{res}}{FSR}$, $\lambda_{res}$ represents the resonance wavelength, FSR is the free spectral range, and $L$ points to the length of the OPCM.[55] As a result, we obtain $\Delta n_{eff} = 0.02$ from Si-doped device while $\Delta n_{eff} = 0.08$ is observed in pure Sb$_2$Se$_3$. Given that we have identified the refractive index at 1588nm for the as-deposited film as 2.7, the refractive index for crystalline films can be estimated from the measured value of $\Delta n_{eff}$, 0.02. To do this, we performed Finite Element Method simulations using Lumerical Mode, and we found an index of 2.85 for crystalline material, which aligns with the results shown **in Figure 1d**. The mapping of the $\Delta n_{eff}$ and sweeping of the refractive index to find the correct experimental value is shown in the Supplementary Figure S7.

Lastly, the Raman spectrum in **Figure 5e** shows the results for crystallized and then re-amorphized material, displaying Si-Se crystalline resonance bonds, which suggests a potential second phase featuring Si-Se bonds always exists, most likely in the form of $SiSe_2$, which is stable an ambient conditions.

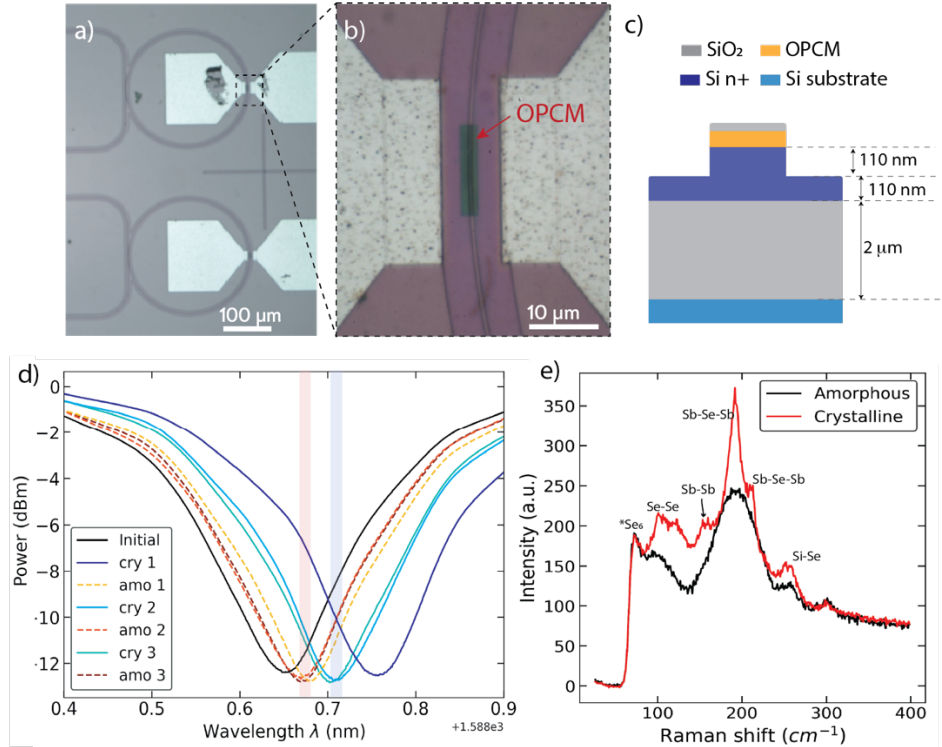

**Figure 5. Reversible switching of Si-doped $Sb_2Se_3$ on PIC.** a) Top-view image of micro-ring resonators with integrated doped-silicon microheaters. b) Zoom-in image a), silver regions are contacts of microheaters, and the green region is the OPCM. c) Cross-section structure of waveguides. d) Resonance shifts around ~1588 nm. The solid lines represent the transmission measured after applying crystallization pulses, while the dashed lines represent transmissions after amorphization pulses. The highlighted regions in red and blue represent the final stable crystalline and amorphous states, respectively. The numbers represent the cycling order. e) Raman spectroscopy measured on the device after the first cycle, i.e., the first re-amorphization and re-crystallization state.

*Table 2.* Parameters used for on-chip switching of 12-μm long OPCM on PIC.

| $Si_x (Sb_2Se_3)_{1-x}$ | Voltage (V) | Width (μs) | Power (mW) | Energy (μJ) |
|---|---|---|---|---|
| x=0% cry | 3.00 | 50 | 25.00 | 2.25 |
| x=0% amo | 7.00 | 0.5 | 245.00 | 0.12 |
| | $\Delta n_{eff} = 0.08, \Delta\varphi = \pi$ | | | |
| x=20% cry | 1.85 | $3\times10^2$ | 23.00 | 6984.00 |
| x=20% amo | 4.70 | 1 | 149.00 | 0.15 |
| | $\Delta n_{eff} = 0.02, \Delta\varphi = 0.2\pi$ | | | |

## 3.2 Phase Engineering on Microheaters

Motivated by the results obtained from NanoDSC and DFT calculations, a plausible composition after phase separation is having $SiSe_2$ and $Sb_2Se_3$, the most thermally stable compounds in the ternary diagram (see **Figure 2a)**. With this in mind, we briefly describe in the Supplementary Section S5 the co-sputtering of $(SiSe_2)_{0.8}(Sb_2Se_3)_{0.2}$ thin films on Ti/Pt microheaters. We successfully switched the materials with consistent pulses (2.5 V/ 1 s for crystallization and 4 V/ 10 μs for amorphization), with the corresponding optical contrast seen in Supplementary Figure S8. $SiSe_2$ has strong covalent bonds, with $T_g$ and $T_c$ of 460 °C and 610 °C, respectively[56], which are close to the $T_m$ of pure $Sb_2Se_3$. Therefore, under the same switching conditions used for pure $Sb_2Se_3$, the $SiSe_2$ phase likely remains amorphous and serves as a stable glassy matrix. At the same time, only the $Sb_2Se_3$ nanocrystals undergo reversible phase transitions, thus forming an optical phase change nanocomposite.

## 4. Conclusion

We studied the effects of Si doping into $Sb_2Se_3$ using several thin film characterization techniques and reconfigurable devices. We found that after the first melt-quenching process, the initially uniform film undergoes phase segregation, forming a nanocomposite consisting of crystalline $Sb_2Se_3$ grains embedded in an amorphous $SiSe_2$ matrix. This behavior differs from that of other conventional phase change materials, which typically form multiple crystalline phases.[57,58] For example, $Ge_4Sb_6Te_7$ forms a coherent nanocomposite with FCC-GeSbTe and SbTe coexisting regions,[59–61] which is found to play a significant role in boosting the overall performance of electrical switching. We exploit this phase segregation, which, contrary to being detrimental, allows for the formation of nanocomposites with effective-medium optical properties [62], where the overall optical response arises from the combination of the amorphous matrix and switchable phase-change grains. Our experimental results demonstrate this novel paradigm by achieving a reduction in $Sb_2Se_3$ optical loss in the visible spectrum with increasing Si doping concentration, albeit at the expense of diminishing the refractive index contrast between states. Moreover, using optimal electrical pulses, we demonstrated robust reconfiguration over multiple cycles of Si-incorporated $Sb_2Se_3$ embedded into photonic integrated circuits. Thus, demonstrating that even after phase segregation, the phase changing nature of $Sb_2Se_3$ allows for optical reconfiguration while the amorphous matrix remains essentially unchanged and contributes to the effective medium response.

Doping also impacts thermal behavior. Using a novel NanoDSC technique, We uncovered that Si dopant suppresses crystallization by slowing the kinetics of phase transition. Yet, phase segregation further complicates the system, as nanocomposites can act as thermal barriers[63], while the distribution of grain boundaries significantly influences the switching behavior.[64] Due to the technical challenges of probing nanometer-sized grains and isolating Si% in a Si-rich multi-layered device structure, we could not conclude the exact chemical stoichiometry of the nanocomposite components after phase segregation. However, our results clearly demonstrate that nanocomposites can lead to engineered optical and thermal properties, which are pivotal for OPCMs; thus, opening a path for future OPCM engineering, leading to tailored properties for applications in nonvolatile optical memories, modulators, metasurfaces, and other areas.

## 5. Experimental Section

*Thin-film Synthesis*

We deposited binary thin film composition spreads on 400 μm-thick sapphire substrates using an ultrahigh vacuum (base pressure: 2 x $10^{-8}$ Torr) magnetron sputtering system (AJA Orion-3) at room temperature. We used two approaches for the spread: a continuous and a patterned gradient, as shown in the supplementary **Figures S8 and S9**. The latter employed a silicon mask placed in contact with the substrate to delineate 15 individual compositions evenly with a 4.5 mm separation. High purity Si (99.995%) and $Sb_2Se_3$ (99.999%) targets (Kurt J. Lesker Co.) used to co-sputter in ultrahigh purity Argon (99.9997%, Airgas) at a pressure of $4.6 \times 10^{-3}$ Torr. The thin film spread was deposited over 25 min using 14W and 17W RF power sources for Si and $Sb_2Se_3$, respectively. The co-sputtering results in an intrinsic continuous thickness gradient (120-30 nm). The thin films were capped with 30 nm $SiO_2$ to avoid surface contamination and, in some cases, with 5 nm Au/Pd to avoid charging under electron microscopy (except TEM). To prepare crystalline samples, the thin film spreads were annealed on a hotplate in an $N_2$ glove box for 20 mins at temperatures ranging from 200°C to 350°C. Note that some samples were annealed longer to compare the phase transition kinetics across compositions. Thin film deposition on PIC and NanoDSC sensors were carried out by placing devices at the position where the targeted composition was achieved.

*Device Fabrication and Measurement Methods*

The PIC devices were fabricated on 220-nm silicon-on-insulator (SOI) wafers, following the fabrication process described by Rios *et al*.[53] We used micro-ring resonators with embedded $n^{++}$-dope microheaters. A 30-nm-thick PCM and 10-nm-thick $SiO_2$ were deposited via sputtering, described in Section 4.1. Electron beam lithography (Elionix ELS-G100 system) was used to pattern the photonic circuitry, followed by $CF_4$ reactive ion etching (Trion RIE). The transmission spectrum measurements and hardware for on-chip electrical switching followed the process thoroughly described in Sun *et al*.[4]

Thin-film samples are deposited onto the NanoDSC sensors for calorimetric measurement. The sensors were fabricated at the Cornell Nanoscale Facility using standard micro-electromechanical systems (MEMS) fabrication methods.[65] A 50 nm Pt film, which performs as the heater and thermometer in the calorimetric cell, is sputtered on the low-stress free-standing $SiN_x$ (100 nm) membrane. The membrane is supported by a 500 μm thick silicon frame. The back side of each sensor is patterned with four Pt electrodes (V+, V-, I+, I-) for four-point resistance measurements. Empty sensors are annealed in vacuum ($2\times10^{-7}$ Torr) to minimize resistance drift (<1% per 1,000 pulses to 750 °C). Thin-film samples are deposited on the front side of the sensor using a self-aligned shadow mask to ensure alignment between sample and calorimetric sensing region (0.5 mm ×5 mm). More details on the design and fabrication of the NanoDSC sensors are described in Ref [66]. Differential mode NanoDSC measurements were performed in vacuum ($2\times10^{-7}$ Torr) by applying synchronized direct-current pulses (duration 1-25 ms) to both sample and reference sensors. For each scan, the Joule heating leads to rapid temperature ramping of the sample from room temperature to a maximum temperature in the 100-720 °C range with a controlled heating rate raging from $8\times10^3$ to $1\times10^6$ K/s, followed by a passive cooling down (~$1\times10^4$ K/s) back to room temperature. The sample heat capacity $C_p(T)$ during the heating cycle is calculated based on the voltage difference between the two sensors with necessary baseline corrections, including heat loss and the difference between the sample/reference addenda. NanoDSC data analysis in differential mode is detailed in our previous work. [52,58]

$Sb_2Se_3$ and $SiSe_2$ (target with 99.5% purity from AJA Inc.) thin films are deposited via co-sputtering using 15W and 11W RF power at $4.6 \times 10^{-3}$ Torr. The microheaters are fabricated with 10 nm- Ti and 50nm- Pt buried in a trench made by thermal oxide, etched by fluorine etcher (Oxford Plasmalab System 100) The bowtie microheaters consist of 100 x100 μm² metal contacts and 12x14 μm² bridge, covered with 30-nm-thick $Sb_2Se_3/SiSe_2$ thin-films. The metals are evaporated followed by lift-off, and $Sb_2Se_3/SiSe_2$ is etched with CF$_4$ reactive ion etching (Trion RIE). All the patterning is done by photolithography (Maskless Aligner, Heidelberg). The electrical switching followed the process as that used on PIC platform.

*Characterization Methods*

Chemical composition of Si-Sb-Se thin film library determined by using wavelength dispersive spectroscopy (WDS) in an electron probe microanalyzer (JXA 8900R Microprobe), with an acceleration voltage of 15 kV. Calibration was done using polished pure metal with an experimental error margin of <0.3 at. %. To capture morphologies and qualitative chemical composition on device, a Tescan GAIA system is used for scanning electron microscopy (SEM) and energy dispersive spectroscopy (EDS) with 10 kV acceleration voltage.

Ellipsometry was performed using a Woollam Variable Angle Spectroscopic Ellipsometry (VASE) system to collect and analyze the optical properties in the 245 -1689 nm spectrum, at 55°, 60° and 65° angles of incidence. The dielectric response fitting was based on Tauc-Lorentz dispersion oscillators and mean squared error (MSE) was used as an index for fitting.

The thin film spread was cut into smaller pieces with the desired compositional range. A Discover powder diffractometer (Bruker C2/D8) of CuK$\alpha$ radiation with a high-temperature stage was used to collect X-ray diffraction (XRD) images. The exposure time was 5 minutes for each frame. The diffraction was integrated into 1D data with the 2θ range from 14° to 60°. To avoid oxidation, the stage was covered by a graphite dome under vacuum (~ 5 Torr). Temperature dependent XRD was conducted from room temperature (30°C) to temperature over T$_m$ (800°C) in discrete steps of 50°C or 100°C. The temperature ramp rate was 20°C/ min with a 5 minutes hold time to reach equilibrium.

The OPCM phase analysis was performed using surface-enhanced Raman spectroscopy (Yvon Jobin LabRam ARAMIS) with a 532 nm laser, a 2400 lines/mm grating and a 100× long-working-distance objective lens.

The 100 nm SiN$_x$ membrane of the NanoDSC sensor makes it possible to conduct direct TEM characterization on the post-scanned thin-film sample using a home-built holder. In this study, sample morphology after NanoDSC scans are obtained using a JEOL 2100 TEM at 200 kV acceleration voltage. The local structure is studied using selected area diffraction with a camera length of 25 cm.

Optical measurements were conducted by directly capturing the intensity of light reflected from the surface of the fabricated device, which was mounted on a motorized XYZ-translation stage. A visible-range LED source (HLV3-24SW) was focused onto the device using a Navita coaxial vision system with up to 30× magnification. To simultaneously visualize the device and acquire the reflected signal, a beam splitter was employed to evenly direct the light to both a CCD camera (Amscope MU500-HS) for imaging and a Horiba spectrometer (Horiba iHR-320) for reflection spectrum acquisition. The measured reflection spectra were normalized by dividing the collected signal by the intensity of a reference beam reflected from a smooth Ti (10 nm)/Pt (90 nm) surface with the same illumination spot size.

*Computational Methods*

The density functional theory (DFT) calculations were conducted using the Vienna Ab initio Simulation Package (VASP)[67] with the Perdew-Burke-Ernzerhof exchange-correlation functional[68]. Our supercell model of Si doped Sb$_2$Se$_3$ follows previous computational studies of the Ge – Sb – Te system[69]; as a cation dopant, Si would occupy Sb sites and introduce vacancies into the cation sublattice. Raman spectroscopy results suggest the presence of Si – Se bonds supporting this assumption (see Results section for details). Taking Si to have its standard 4+ oxidation state, we used the following defect reaction was used:

$$3SiSe_2 \xrightarrow[Sb_2Se_3]{} 3Si_{Sb}^{\circ} + V_{Sb}''' + 6Se_{Se}$$

We created 4 supercells of orthorhombic Sb$_2$Sb$_3$ with increasing Si concentrations: 0% Si (0 Si, 24 Sb, 36 Se), 5.1% Si (3 Si, 20 Sb, 36 Se), 10.3% Si (6 Si, 16 Sb, 36 Se), and 15.8% Si, (9 Si, 12 Sb, 36 Se). To determine the optimal arrangement of Si atoms and Sb vacancies, we used pymatgen[70] to enumerate every possible arrangement of Si atoms and vacancies in each supercell. Then, the 10 structures with the lowest Ewald energy[71] were relaxed with DFT. The structures with the lowest DFT energy were selected as the representative structures for subsequent band gap calculations. The band gap calculations were performed using the parameters generated by pymatgen[70] as in the Materials Project[72].

The simulations of the silicon waveguide modes for the different phase states of Si-Sb$_2$Se$_3$ were performed using Lumerical MODE (Ansys®).

## Acknowledgements

C.A.R.O., J.H., N.Y, and I.T. acknowledge funding provided by the National Science Foundation (awards ECCS-2210168/2210169, ECCS-2430920, and DMR-2329087/2329088), supported in part by industry partners, as specified in the Future of Semiconductors (FuSe) program. J. Z. and L. H. A. acknowledge the financial support by NSF-DMR-1409953 and NSF-DMR-1809573. The NanoDSC sensors were fabricated at the Cornell Nanoscale Facility (project #522-94), a member of the National Nanotechnology Infrastructure Network (NNIN). Materials characterization was conducted in part at the Materials Research Laboratory, University of Illinois, and the NanoCenter, University of Maryland.